\documentclass[twocolumn]{aastex63}

\usepackage{amsmath}
\usepackage{array}
\usepackage{enumitem}
\newcolumntype{C}[1]{>{\centering\arraybackslash}p{#1}}
\newcolumntype{L}[1]{>{\raggedright\let\newline\\\arraybackslash\hspace{0pt}}m{#1}}

\submitjournal{ApJL}

\shorttitle{$^{22}$Ne Phase Separation in WDs}
\shortauthors{Blouin et al.}

\begin{document}

\title{$^{22}$Ne Phase Separation As A Solution To The Ultramassive White Dwarf Cooling Anomaly}

\correspondingauthor{Simon Blouin}
\email{sblouin@lanl.gov}

\author[0000-0002-9632-1436]{Simon Blouin}
\affiliation{Los Alamos National Laboratory, PO Box 1663, Los Alamos, NM 87545, USA}

\author[0000-0002-8844-6124]{J{\'e}r{\^o}me Daligault}
\affiliation{Los Alamos National Laboratory, PO Box 1663, Los Alamos, NM 87545, USA}

\author[0000-0001-6800-3505]{Didier Saumon}
\affiliation{Los Alamos National Laboratory, PO Box 1663, Los Alamos, NM 87545, USA}

\begin{abstract}
The precise astrometric measurements of the Gaia Data Release~2 have opened the door to detailed tests of the predictions of white dwarf cooling models. Significant discrepancies between theory and observations have been identified, the most striking affecting ultramassive white dwarfs. \cite{cheng2019} found that a small fraction of white dwarfs on the so-called Q branch must experience an extra cooling delay of $\sim 8\,{\rm Gyr}$ not predicted by current models. $^{22}$Ne phase separation in a crystallizing C/O white dwarf can lead to a distillation process that efficiently transports $^{22}$Ne toward its center, thereby releasing a considerable amount of gravitational energy. Using state-of-the-art Monte Carlo simulations, we show that this mechanism can largely resolve the ultramassive cooling anomaly if the delayed population consists of white dwarfs with moderately above-average $^{22}$Ne abundances. We also argue that $^{22}$Ne phase separation can account for the smaller cooling delay currently missing for models of white dwarfs with more standard compositions.
\end{abstract}
\keywords{Cosmochronology --- Degenerate matter --- Plasma physics --- Stellar evolution --- Stellar interiors --- White dwarf stars}

\section{Introduction}
\label{sec:intro}
White dwarf evolution is often depicted as a simple, uneventful cooling process, implying that they naturally lead to accurate age determinations. White dwarfs are indeed very useful cosmic clocks. They have been used to determine the ages of individual stellar populations \citep{hansen2007} and to reconstruct the star formation history of the Milky Way \citep{tremblay2014,kilic2017,fantin2019}. When paired with a stellar companion, they also become useful benchmarks for harder-to-model objects \citep[e.g., M, L, and T dwarfs,][]{lam2020,meisner2020}. They could even prove useful to track the galactic evolution of lithium \citep{kaiser2021}.

However, some aspects of white dwarf cooling remain poorly understood, meaning that the theoretical cooling tracks routinely used to infer white dwarf ages may not be as accurate as they ought to be. The existence of those inaccuracies has been clearly demonstrated during the last few years thanks to the Gaia Data Release~2 \citep{gaiadr2a,gaiadr2b}. \cite{cheng2019} revealed the most salient example of the shortcomings of current evolution models by showing that some ultramassive white dwarfs ($M_{\star} \gtrsim 1.05\,M_{\odot}$) experience an additional cooling delay of $\sim 8\,{\rm Gyr}$ compared to theoretical predictions. Those objects are found in a region of the Gaia color--magnitude diagram (CMD) known as the Q branch, which corresponds to an overdensity of objects that coincides with the high-mass tail of core crystallization \citep{tremblay2019}.

White dwarfs experiencing this additional cooling delay most likely have C/O cores, as the C/O crystallization sequence (and not the O/Ne crystallization sequence) is consistent with the location of the observed overdensity in the Gaia CMD \citep[][Figure~3]{bauer2020}. In apparent contradiction with this observation, standard single-star evolution models predict the formation of O/Ne cores in ultramassive white dwarfs \citep{siess2007}, and ultramassive white dwarfs formed from the merger of two C/O white dwarfs are also predicted to have O/Ne cores \citep{schwab2021}. However, \cite{althaus2021} recently proposed two single-star evolution scenarios that can lead to the formation of ultramassive C/O white dwarfs.

Assuming a mass fraction of $^{22}$Ne $X(^{22}{\rm Ne})=0.02$\footnote{We use $X$ for mass fractions and $x$ for number fractions.}, \cite{cheng2019} note that, if released during crystallization, the gravitational energy of $^{22}$Ne stored in ultramassive C/O white dwarfs is sufficient to generate a $\sim 6 - 9\,{\rm Gyr}$ cooling delay. Such a large effect is possible owing to the neutron-rich nature of $^{22}$Ne ($A > 2Z$). The most commonly discussed mechanism to release this potential energy is the gravitational settling of $^{22}$Ne in the liquid phase \citep{bildsten2001}. However, current models predict that simple gravitational settling is not efficient enough to give rise to multi-Gyr cooling delays \citep[e.g.,][]{bauer2020}, in part because crystallization of the C/O core strongly inhibits the extent of the diffusion of $^{22}$Ne in the liquid phase.

\cite{camisassa2020} proposed that the objects undergoing this extra delay are extremely rich in $^{22}$Ne, with a $^{22}$Ne mass fraction $X(^{22}{\rm Ne})=0.06$ instead of the $X(^{22}{\rm Ne})=0.014$ value expected for white dwarfs issued from the single-star evolution of solar-metallicity progenitors \citep{cheng2019}. While quadrupling $X(^{22}{\rm Ne})$ can indeed produce a multi-Gyr cooling delay, more work is needed to explain how ultramassive white dwarfs could acquire such a high $^{22}$Ne abundance. Simulations by \cite{staff2012} show that mergers of He and C/O white dwarfs can lead to a sizeable $^{22}$Ne enrichment. Yet, the stars produced in those merger simulations had a mass of only $0.9\,M_{\odot}$, meaning that those results may not be directly applicable to ultramassive white dwarfs. Another avenue, explored by \cite{bauer2020}, is to increase the rate of $^{22}$Ne gravitational settling through the formation of solid clusters containing a few thousand $^{22}$Ne ions. However, subsequent molecular dynamics simulations have shown that the formation of such clusters cannot take place at the low $X(^{22}{\rm Ne})$ found in C/O white dwarfs \citep{caplan2020}. Other avenues unrelated to $^{22}$Ne sedimentation have also been explored, but cannot explain the missing multi-Gyr delay \citep{horowitz2020}.

In this letter, we accurately calculate the melting curve of C/O/Ne mixtures to show that the gravitational energy of $^{22}$Ne stored in ultramassive white dwarfs can be efficiently released through a {\it phase separation} process. Assuming a modest $^{22}$Ne enrichment (half of that assumed by \citealt{camisassa2020}), this mechanism leads to multi-Gyr cooling delays that can largely explain the ultramassive white dwarf cooling anomaly. We also argue that this process can solve a separate, smaller cooling delay problem that affects standard-composition white dwarfs.

\newpage

\section{The Phase Separation of Neon--22}
\label{sec:ps}
The possibility of $^{22}$Ne phase separation\footnote{By ``$^{22}$Ne phase separation'' we mean the fractionation of $^{22}$Ne, whereby the $^{22}$Ne concentrations in the coexisting liquid and solid phases differ.} and its effects on the cooling of C/O white dwarfs were first discussed by \cite{isern1991}. The general idea is that depending on the composition of the C/O/Ne plasma and the exact shape of the C/O/Ne phase diagram, the solid crystals that are formed when the crystallization temperature is reached can be {\it depleted} in $^{22}$Ne with respect to the liquid mixture. If the $^{22}$Ne depletion is large enough, those crystals are lighter than the surrounding liquid and float upward, away from the crystallization front. The rising crystals eventually melt in lower density regions where their constituent ions are mixed via Rayleigh--Taylor instabilities. This is analogous to a distillation\footnote{``Creaming'' may be a better term to describe this process, but we use ``distillation'' to remain consistent with previous work on this topic.} process and it gradually displaces $^{22}$Ne-rich liquid downward. The $^{22}$Ne abundance in the liquid at the liquid--solid interface thereby increases until it reaches a critical value for which the C/O/Ne phase diagram predicts that there is no more phase separation (i.e., when the liquidus and solidus meet). The plasma then freezes at this constant composition, forming solid $^{22}$Ne-rich layers. The crystallization of the remaining liquid C/O mixture (now free of $^{22}$Ne) then continues as usual.

The amount of gravitational energy released in this process critically depends on the exact shape of the C/O/Ne phase diagram. If the distillation process starts immediately at the beginning of the crystallization of the C/O core (i.e., when the core is still fully liquid), then a very large cooling delay is possible. In this scenario, favored by the exploratory calculations of \cite{isern1991} and \cite{segretain1994}, all the $^{22}$Ne from the liquid core is transported to the center of the white dwarf where it forms a compact $^{22}$Ne-rich central core. It is also possible, as suggested by the C/O/Ne phase diagram of \cite{segretain1996}, that the distillation process only starts later, when a fraction of the core is already solidified. A $^{22}$Ne-rich shell is then formed around a central core with a nearly unperturbed $^{22}$Ne concentration. In this scenario, a smaller amount of $^{22}$Ne is transported downward and not as deep in the gravitational well of the star, with a more modest energy release.

\cite{caplan2020} recently used the semi-analytic approach of \cite{medin2010} to map the C/O/Ne phase diagram.\footnote{\cite{hughto2012} also studied the liquid--solid coexistence of C/O/Ne mixtures using two-phase molecular dynamics simulations. However, only a handful of compositions were simulated, so that not much can be said about the phase separation of  $^{22}$Ne from their results.} While they do not discuss this possibility, we note that their Figure~4 confirms that the solid phase can be depleted in $^{22}$Ne when the $^{22}$Ne concentration remains small in the liquid phase, enabling the distillation process described above. We also remark that their phase diagram indicates that $^{22}$Ne distillation will continue (the $^{22}$Ne concentration in the liquid will remain higher than that in the solid, $x_{\rm Ne}^{\ell}> x_{\rm Ne}^{s}$)\footnote{The solid must be less dense than the liquid for distillation to take place, which is almost (but not rigorously, see Section~\ref{sec:mc}) equivalent to the $x_{\rm Ne}^{\ell}> x_{\rm Ne}^{s}$ condition.} until $(x_{\rm C},x_{\rm O},x_{\rm Ne}) \simeq (0.8,0.0,0.2)$. This composition corresponds to where the liquidus and solidus meet in the two-component C/Ne phase diagram (see also Figure~4 of \citealt{medin2010} and Figure~5 of \citealt{ogata1993}).

While past calculations of the C/O/Ne phase diagram support the existence of $^{22}$Ne distillation, they are not precise enough to allow an accurate quantitative assessment of the impact of this process on white dwarf cooling. In particular, a high-resolution version of the C/O/Ne phase diagram at small $^{22}$Ne concentrations is required. The exact shape of the phase diagram in this region is critical, as it dictates even the qualitative outcome of the distillation process (the formation of a $^{22}$Ne-rich central core vs a $^{22}$Ne-rich shell closer to the surface).

\section{Monte Carlo Simulations}
\label{sec:mc}
To obtain a high-resolution, high-precision version of the C/O/Ne phase diagram at small $^{22}$Ne concentrations, we turn to the Clapeyron integration technique that we have recently developed to map the phase diagrams of dense plasmas (Blouin \& Daligault 2021, under review) and applied to the two-component C/O plasma \citep{blouin2020}. Briefly, this method consists of directly integrating the liquid--solid coexistence line using the appropriate Clapeyron equation for phase transitions at constant temperature and pressure (here, Equation~A2 of Blouin \& Daligault 2021), which we evaluate using Monte Carlo simulations where the full electron--ion plasma is considered. All our simulations use $N=686$ ions, which is enough to mitigate finite-size effects even at very small concentrations ($x_{\rm Ne} \sim 0.002$). We verified this point by performing simulations with up to $N=4000$ ions. Each simulation is executed for $7 \times 10^6$ Monte Carlo iterations, of which the last $5 \times 10^6$ are used to evaluate the average thermodynamics quantities needed to integrate the Clapeyron equation. We assume a fixed $P=10^{24}\,{\rm dyn\,cm}^{-2}$ pressure, a typical value for white dwarf cores. We integrate the melting line at constant temperature $T$, meaning that many distinct integrations are needed to fully map the C/O/Ne phase diagram in the three-dimensional composition--temperature space (we use 11 different temperatures). It is useful to express the temperature as
\begin{equation}
\Gamma_{{\rm C}} = \frac{e^2}{a_e k_B T} Z_{\rm C}^{5/3},
\end{equation}
where $e$ is the elementary charge, $a_e= \left(3/4 \pi n_e\right)^{1/3}$ with $n_e$ the electron density, $k_B$ is Boltzmann's constant, and $Z_{\rm C}=6$. At constant $P$, $n_e$ is essentially constant in dense degenerate plasmas and $\Gamma_{{\rm C}}$ is therefore a useful dimensionless measure of $T$. For reference, a pure C plasma solidifies at $\Gamma_{{\rm C}} \simeq 175$. The 11 temperatures we simulate are equivalent to $\Gamma_{\rm C}$ ranging from 132 to 182. Those values in turn correspond to the melting temperatures, $\Gamma_{{\rm C},m}$, of C/O plasmas with $x_{\rm O}^{\ell}$ ranging from 0.28 to 0.79.

\begin{figure}
    \includegraphics[width=\columnwidth]{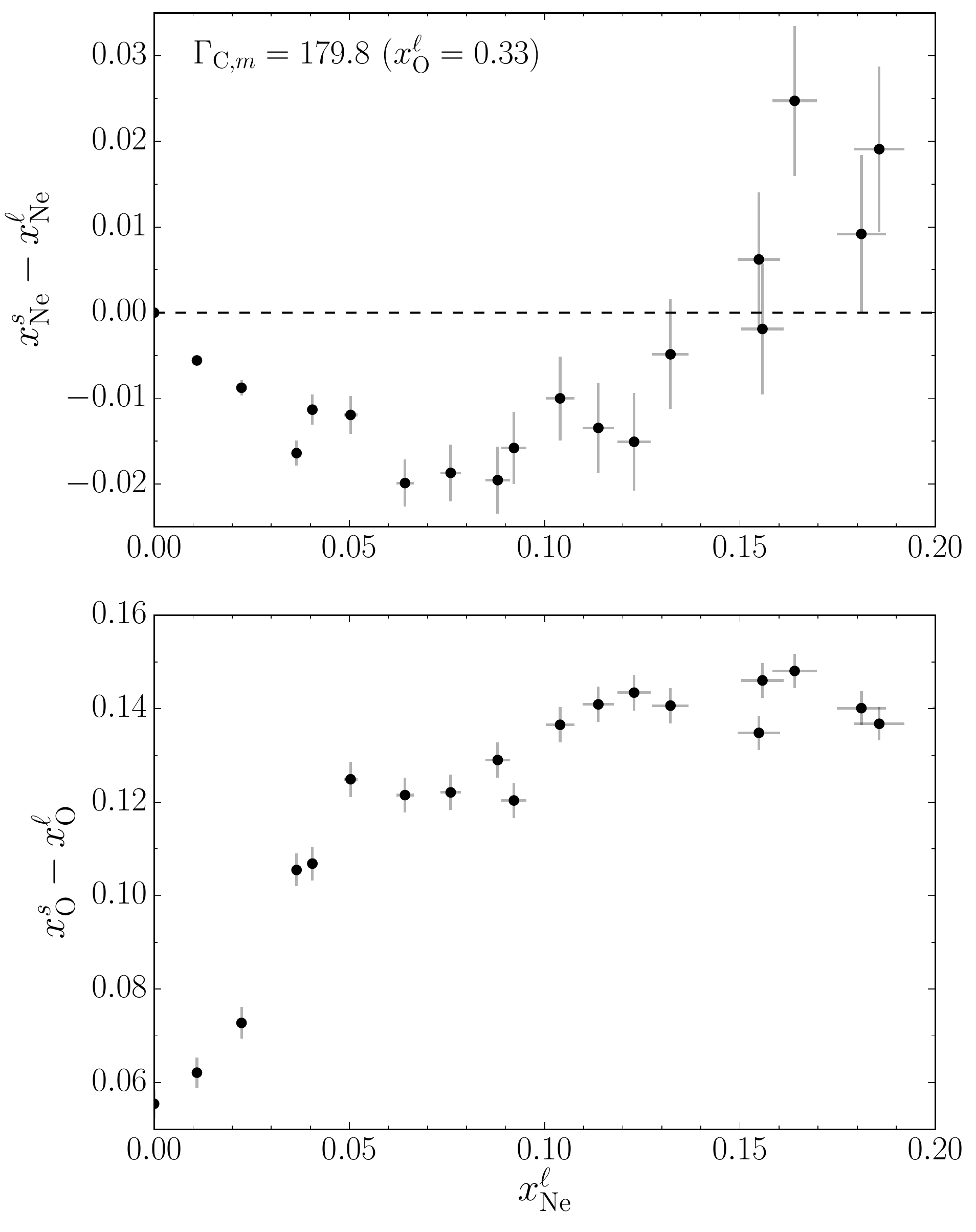}
    \caption{Ne and O concentration changes at the liquid--solid phase transition as a function of the Ne concentration in the liquid phase for a C/O/Ne mixture with $\Gamma_{{\rm C},m}=179.8$. The error bars were obtained by applying the block-averaging technique to our Monte Carlo trajectories and correspond to $1\sigma$ confidence intervals. $x_{\rm O}^{\ell}=0.33$ is the O concentration at $x_{\rm Ne}^{\ell}=0$ on the coexistence line.}
  \label{fig:deltax}
\end{figure}

Figure~\ref{fig:deltax} shows the result of a low-resolution integration of the phase diagram at $\Gamma_{\rm C}=179.8$ (the melting temperature of a C/O plasma with $x_{\rm O}^{\ell}=0.33$) between $x_{\rm Ne}^{\ell}=0.0$ and $x_{\rm Ne}^{\ell}=0.2$. As with the two-component C/O phase diagram, we find that the solid is enriched in O compared to the liquid. As for Ne, the solid can either be enriched or depleted with respect to the liquid depending on the Ne abundance in the liquid. For the temperature used in Figure~\ref{fig:deltax}, the solid is depleted in Ne as long as $x_{\rm Ne}^{\ell} \lesssim 0.15$. It is in this regime that $^{22}$Ne distillation is possible.

\begin{figure}
    \includegraphics[width=\columnwidth]{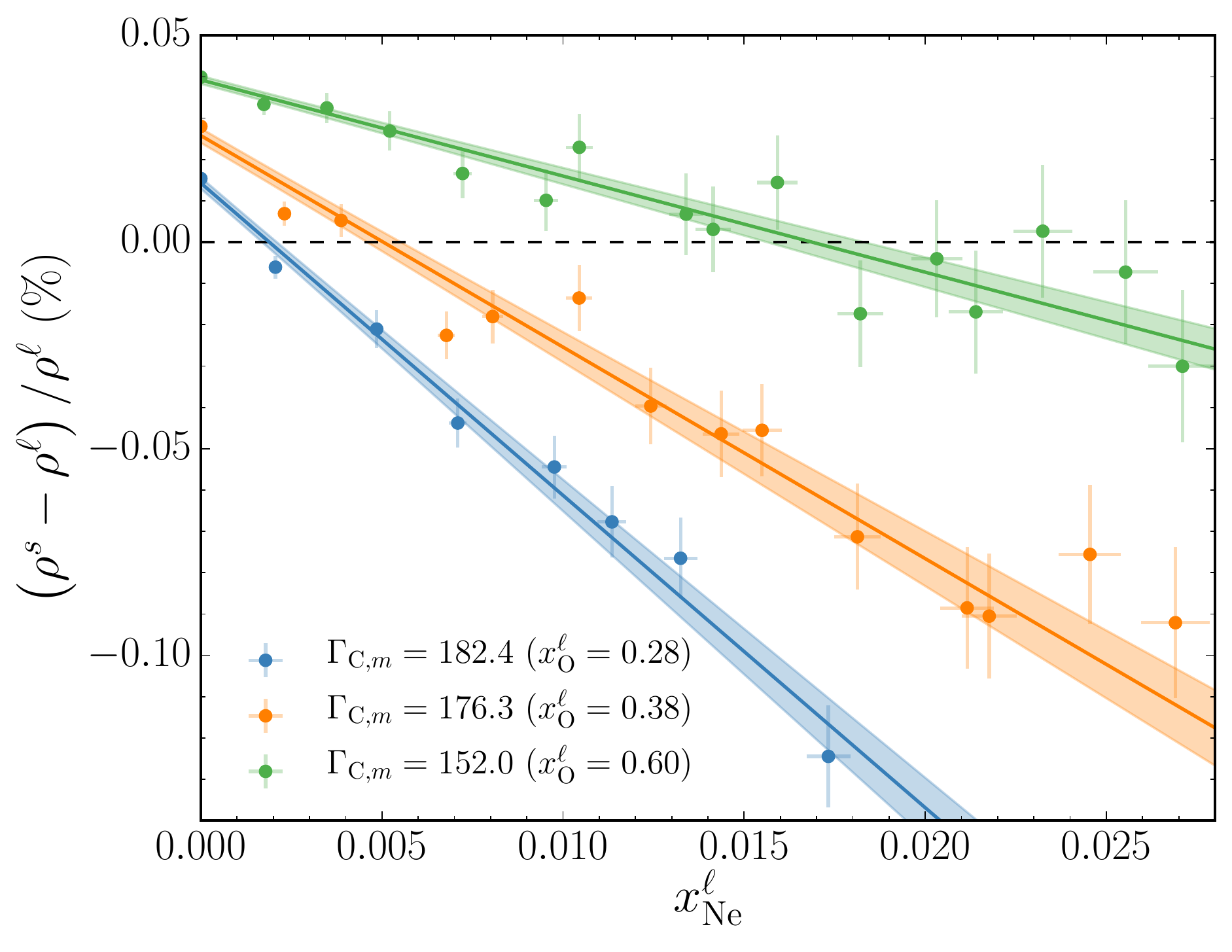}
    \caption{Relative density differences between the liquid and solid phases at the phase transition. The three colors correspond to three different temperatures, as indicated in the legend. As in Figure~\ref{fig:deltax}, the $x_{\rm O}^{\ell}$'s correspond to the O concentrations in the liquid phase on the coexistence line at $x_{\rm Ne}^{\ell}=0$. The lines are linear fits to our simulation results (shown as circles with error bars) and the shaded regions around them indicate the fit uncertainty.}
  \label{fig:deltarho}
\end{figure}

To precisely determine when the distillation process begins, we performed a set of high-resolution integrations of the phase diagram between $x_{\rm Ne}^{\ell}=0.00$ and $x_{\rm Ne}^{\ell}=0.05$. For each point along the coexistence line, we evaluate the mass densities $\rho$ of both phases. Figure~\ref{fig:deltarho} shows how the solid--liquid density difference varies as a function of $x_{\rm Ne}^{\ell}$ for 3 of the 11 temperatures for which we have integrated the C/O/Ne phase diagram. Temperatures and $^{22}$Ne concentrations that are below the $\rho^{s}=\rho^{\ell}$ line correspond to conditions where $^{22}$Ne distillation operates. This regime is reached at low melting temperatures (corresponding to a high $\Gamma_{{\rm C},m}$ and a low O concentration) and/or high $^{22}$Ne concentrations. Note that even though we found that the solid is always depleted in $^{22}$Ne in the $x_{\rm Ne}^{\ell} \rightarrow 0$ limit, its density is not always smaller than that of the liquid as its O enrichment must also be considered.

\newpage

\section{Implications for White Dwarf Cooling}
\label{sec:wd}
\begin{figure*}
\centering
    \includegraphics[width=1.65\columnwidth]{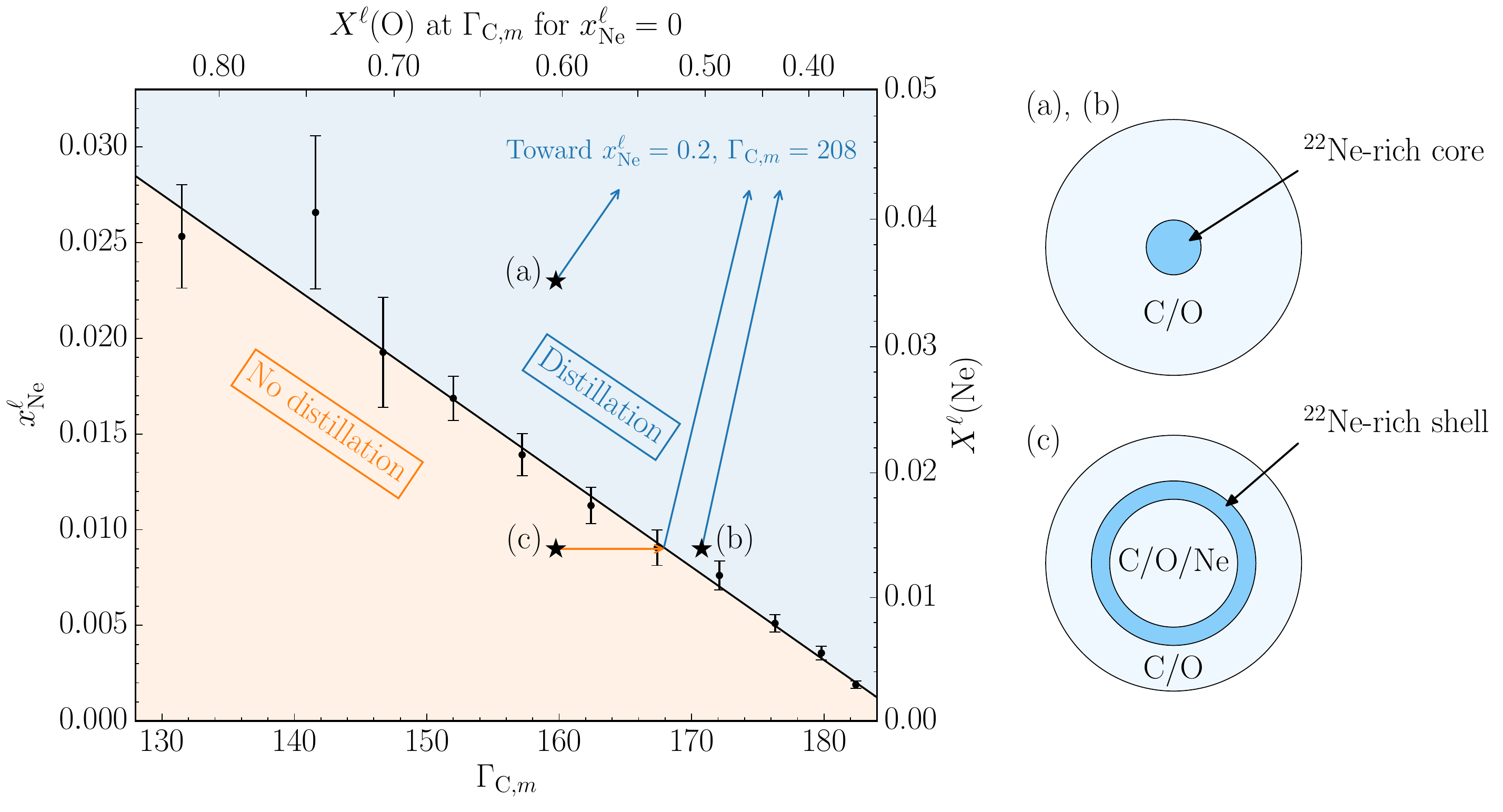}
    \caption{The circles with error bars indicate the conditions where we find that $\rho^s = \rho^{\ell}$ at the phase transition. The orange region below the line formed by those error bars corresponds to the regime where the solid sinks and no distillation takes place. The blue region corresponds to where the solid is lighter than the liquid, leading to $^{22}$Ne distillation. The top horizontal axis gives the O mass fraction in the liquid for a C/O plasma that crystallizes at the temperature given by the bottom axis. The different scenarios (a), (b), and (c) are discussed in the text.}
  \label{fig:master}
\end{figure*}

Figure~\ref{fig:master} translates the results of Figure~\ref{fig:deltarho} in the $(\Gamma_{\rm C},x_{\rm Ne}^{\ell})$ plane for all 11 simulated temperatures. A key finding is that {\it both} scenarios described in Section~\ref{sec:ps} (the formation of a $^{22}$Ne-rich central core or shell) are possible depending on the white dwarf's initial composition. 

If the central composition of the white dwarf is rich enough in $^{22}$Ne (scenario (a) in Figure~\ref{fig:master}), distillation will start directly at the onset of crystallization when the melting temperature $\Gamma_{{\rm C},m}$ is reached in the central layers. The core will remain entirely liquid until the distillation process described in Section~\ref{sec:ps} increases the $^{22}$Ne concentration in the central layers to the critical $x_{\rm Ne}=0.2$ value (the solid crystals keep floating upward, so no solid core can be formed). This $^{22}$Ne enrichment process is symbolized by the blue arrow in Figure~\ref{fig:master}. The central layers will then freeze at the composition $(x_{\rm C},x_{\rm O},x_{\rm Ne}) = (0.8,0.0,0.2)$, which is reached at $\Gamma_{{\rm C},m}=208$. After that, the crystallization of the remaining C/O core will proceed as usual. This will lead to the formation of a $^{22}$Ne-rich C/Ne solid core surrounded by a C/O mantle completely depleted of $^{22}$Ne. This final state can also be attained if the central mixture is O poor, in which case the melting temperature is lower and distillation can start right at the beginning of the crystallization process even with a small $^{22}$Ne concentration. This corresponds to scenario (b) in Figure~\ref{fig:master}. 

If a more standard composition is assumed (e.g., scenario (c), $X({\rm O})=0.60$ and $X(^{22}{\rm Ne})=0.014$, or $x_{\rm O}=0.53$ and $x_{\rm Ne}=0.009$), then no $^{22}$Ne distillation can initially occur and crystallization takes place as in the case of a two-component C/O plasma without any significant change to the $^{22}$Ne distribution. As the crystallization front progresses outward, the liquid is gradually depleted in O due to C/O phase separation and it freezes at increasingly lower temperatures \citep{horowitz2010,althaus2012,blouin2020}. Eventually, $\Gamma_{{\rm C},m}$ crosses the boundary that delimits the distillation regime and $^{22}$Ne distillation can start (this corresponds to the tip of the orange arrow in Figure~\ref{fig:master}). The progression of the crystallization front is halted until the distillation process increases the $^{22}$Ne abundance in the liquid at the liquid--solid interface to $x_{\rm Ne}=0.2$. The $^{22}$Ne-rich layers surrounding the solid core then freeze, and the crystallization of the remaining C/O layers continues following the two-component C/O phase diagram. This leads to a final structure where a C/O/Ne central core is surrounded by a $^{22}$Ne-rich C/Ne shell, which is itself surrounded by a C/O outer shell completely depleted of $^{22}$Ne.

How much does this process affect white dwarf cooling? To answer this question, we generate white dwarf structures using STELUM (\citealt{bedard2020}; Bédard et al. 2021, in preparation) and compute their binding energies
\begin{equation}
B = - \left( \int_0^{M_{\star}} u dm - \int_0^{M_{\star}} G \frac{m}{r} dm \right),
\end{equation}
where $u$ is the internal energy per unit mass, $G$ is the gravitational constant, and the integrals are performed over the whole star. The net energy released by the distillation process corresponds to the difference in binding energy, $\Delta B$, between the white dwarf structure at the onset of $^{22}$Ne phase separation and the structure obtained once it ends. 

For the initial structure, we assume that $^{22}$Ne is homogenously mixed throughout the C/O core. This is not strictly correct as $^{22}$Ne gravitational settling in the liquid phase will create a $^{22}$Ne concentration gradient, but this effect is small \citep[e.g., Figures~2 and~3 of][]{garcia2008} and can be neglected for our purpose. For simplicity, we also assume that the C/O abundance ratio is constant throughout the core. For the final structure, we place all the $^{22}$Ne in a central core of mass $M_{\star}X(^{22}{\rm Ne})/X^a(^{22}{\rm Ne})$, where $M_{\star}$ is the white dwarf's mass, $X(^{22}{\rm Ne})$ is its overall $^{22}$Ne mass fraction, and $X^a(^{22}{\rm Ne})=0.3$ is the $^{22}$Ne mass fraction in the central core (i.e., $x_{\rm Ne}=0.2$). If a $^{22}$Ne-rich shell is formed instead, then its mass is given by $M_{\star}^{\ell} X(^{22}{\rm Ne})/X^a(^{22}{\rm Ne})$, where $M_{\star}^{\ell}$ is the mass of the fraction of the core that is still liquid when the distillation process starts. Table~\ref{tab:delay} gives the resulting $\Delta B$ for different scenarios. We also estimate the resulting cooling delay by dividing this $\Delta B$ by the average luminosity of the star over the distillation process.\footnote{A thick $M_{\rm H}/M_{\star} = 10^{-4}$ H envelope is assumed in all cases.} 

\begin{table}
\centering
 \caption{Effect of $^{22}$Ne phase separation. \label{tab:delay}}
\begin{tabular}{cccccc}
\hline
\hline
$M_{\star}$ & $X({\rm O})$ & $X(^{22}{\rm Ne})$ & $\log L/L_{\odot}$$^a$ & $\Delta B$ & $\Delta \tau$ \\
($M_{\odot}$) & & & & $(10^{47}\,{\rm erg})$ & (Gyr) \\
\hline
\multicolumn{6}{c}{Formation of a $^{22}$Ne-rich central core}\\
\hline
1.0 & 0.50 & 0.035 & $-3.0$ & 10 & 9.0 \\
1.0 & 0.60 & 0.035 & $-2.9$ & 10 & 6.4 \\
1.1 & 0.50 & 0.035 & $-2.8$ & 15 & 7.1 \\
1.1 & 0.60 & 0.035 & $-2.6$ & 15 & 5.0 \\
1.2 & 0.50 & 0.035 & $-2.4$ & 23 & 4.6 \\
1.2 & 0.60 & 0.035 & $-2.3$ & 23 & 3.6 \\
\hline
\multicolumn{6}{c}{Formation of a $^{22}$Ne-rich shell}\\
\hline
0.6 & 0.60 & 0.014 & $-4.1$ & 0.18 & 1.8 \\
0.8 & 0.60 & 0.014 & $-3.8$ & 0.34 & 2.0 \\
1.0 & 0.60 & 0.014 & $-3.5$ & 0.66 & 1.6 \\
\hline
\multicolumn{6}{L{8.2cm}}{$^a$Average luminosity of the star over the distillation process.}
\end{tabular}
\end{table}

The first portion of Table~\ref{tab:delay} gives the impact of $^{22}$Ne phase separation on ultramassive $^{22}$Ne-rich white dwarfs that will form a $^{22}$Ne-rich central core as in scenario (a) of Figure~\ref{fig:master}. For those models, we use the $X(^{22}{\rm Ne})=0.035$ value discussed in \cite{bauer2020} for white dwarfs that descend from stars formed in $\alpha$-rich environments. However, our conclusions are not limited to this particular scenario; any other scenario that leads to a moderate $^{22}$Ne enrichment can be envisaged \citep[e.g., the merger of two white dwarfs,][]{blouin2020,camisassa2020}. We find that $^{22}$Ne phase separation induces an additional $4-9\,{\rm Gyr}$ cooling delay for those objects, which explains most of the ultramassive white dwarf cooling anomaly. An abundance of $X(^{22}{\rm Ne})=0.035$ is expected to be rare in white dwarfs, which is consistent with the fact that only $\sim 6$\% of ultramassive white dwarfs are affected by the cooling anomaly \citep{cheng2019}. Depending on the mass and composition of the white dwarf, the delay induced by $^{22}$Ne phase separation can be short of the 8~Gyr delay inferred by \cite{cheng2019}, but there are additional factors to consider:
\begin{enumerate}[wide, labelwidth=!, labelindent=0pt]
\item The $^{22}$Ne mass fraction for the delayed ultramassive white dwarfs may well be higher than the $X(^{22}{\rm Ne})=0.035$ value assumed here.
\item As explained in \cite{bauer2020}, the missing cooling delay can be reduced to $\sim 6\,$Gyr by changing some assumptions in \citeauthor{cheng2019}'s analysis, such as the thick disk age, the age--velocity dispersion relation, and the star formation history.
\item The C/O white dwarf cooling tracks used by \cite{cheng2019} for $M_{\star} < 1.10\,M_{\odot}$ do not include C/O phase separation \citep{fontaine2001}, meaning that they underestimate the time those objects stay on the Q branch by up to $\sim 1\,$Gyr \citep{blouin2020}.
\item \cite{cheng2019} use O/Ne white dwarf cooling tracks \citep{camisassa2019} for $M_{\star} \geq 1.10\,M_{\odot}$, but it is increasingly clear that the objects that form the delayed population have C/O cores \citep{bauer2020,camisassa2020}. As pointed out by \cite{bauer2020}, O/Ne white dwarfs crystallize much earlier in their evolution than C/O white dwarfs (i.e., at higher luminosities), meaning that the latent heat and gravitational energy released during crystallization induce a shorter cooling delay and that $^{22}$Ne diffusion has less time to operate. From Figure~2 of \cite{camisassa2020}, we infer that this led \cite{cheng2019} to overestimate by $\sim 1.5\,$Gyr the cooling delay experienced by $M_{\star} \geq 1.10\,M_{\odot}$ objects assuming they have C/O cores.
\end{enumerate}
Overall, the cooling delay provided by $^{22}$Ne phase separation appears to be the key piece of missing physics required to explain the ultramassive white dwarf cooling anomaly. Note that our results provide further support for the idea that the delayed population identified by \cite{cheng2019} have C/O cores, as $^{22}$Ne distillation cannot take place in O/Ne cores. Neglecting quantum effects in the ionic degrees of freedom, the phase diagram does not discriminate between $^{20}$Ne and $^{22}$Ne. The total $^{20}{\rm Ne}$+$^{22}{\rm Ne}$ abundance being higher than that of the azeotrope \citep[Figure 1 of][]{camisassa2019}, the solid will be enriched in $^{20}$Ne and $^{22}$Ne, making it denser than the liquid. A sedimentation process analogous to that occurring in C/O white dwarfs will take place.

Any solution to the ultramassive cooling anomaly must also preserve the existing agreement between theory and observations for normal-composition white dwarfs. For instance, the luminosity function of massive white dwarfs in the range $0.9<M_{\star}/M_{\odot}<1.1$ is relatively well fitted using current evolution models \citep{tremblay2019,blouin2020}. An additional multi-Gyr cooling delay for those objects can be ruled out. If a standard $X(^{22}{\rm Ne})=0.014$ abundance is assumed instead of $X(^{22}{\rm Ne})=0.035$, a $^{22}$Ne-rich shell is formed, as in scenario (c) of Figure~\ref{fig:master}. Much less gravitational energy is then released, which, as needed, leads to shorter cooling delays (Table~\ref{tab:delay}). 

Instead of occurring at the onset of crystallization, those delays now occur once $\sim 60$\% of the core is already crystallized (assuming an homogeneous $X({\rm O})=0.60$ C/O initial profile). Very promisingly, a cooling delay of the order of $\sim 1\,$Gyr that manifests itself once $\sim 60$\% of the core is crystallized is missing from current C/O cooling tracks (Figure~2 of \citealt{blouin2020} and Figure~19 of \citealt{kilic2020}). $^{22}$Ne phase separation is very likely the solution to this second problem, although detailed evolutionary calculations that include $^{22}$Ne gravitational settling in the liquid phase will be needed to confirm this hypothesis.

Finally, what about $^{22}$Ne-rich normal-mass white dwarfs? A $0.6\,M_{\odot}$ white dwarf with $X(^{22}{\rm Ne})=0.035$ and $X({\rm O})=0.60$ would experience a very long cooling delay of $\approx 12\,{\rm Gyr}$ ($\Delta B = 2.1 \times 10^{47}\,{\rm erg}$) during the formation of its $^{22}$Ne-rich central core at $\log L/L_{\odot} \approx -3.9$. However, no discrepancy between the dynamical and photometric ages of the sort discovered for ultramassive white dwarfs \citep{cheng2019} has so far been identified for normal-mass objects. If confirmed, this non-detection can be explained in two ways:
\begin{enumerate}[wide, labelwidth=!, labelindent=0pt]
\item Very few normal-mass white dwarfs are enriched in $^{22}$Ne to the level required to form a $^{22}$Ne-rich central core. This would suggest that a negligible number of white dwarfs in the solar neighborhood were formed in $\alpha$-rich environments, challenging \citeauthor{bauer2020}'s hypothesis that the delayed ultramassive population originates from such environments. In this context, the idea that the delayed ultramassive white dwarfs acquired their high $^{22}$Ne content during a merger event \citep{blouin2020,camisassa2020} may be more promising.
\item Alternatively, the central layers of normal-mass white dwarfs may be much more rich in O then currently assumed \citep[e.g., $X({\rm O})\simeq 0.85$,][]{giammichele2018}, which would lead to the formation of a $^{22}$Ne-rich shell (Figure~\ref{fig:master}), with a much smaller cooling delay.
\end{enumerate}

\section{Conclusion}
\label{sec:conclu}
We have shown that ultramassive white dwarfs experiencing a long delayed cooling on the Q branch likely correspond to a population of $^{22}$Ne-rich C/O white dwarfs in which $^{22}$Ne phase separation induces a multi-Gyr cooling delay through the formation of a $^{22}$Ne-rich central core. Such multi-Gyr delays are not observed for the vast majority of white dwarfs since, for canonical element abundances, $^{22}$Ne phase separation leads instead to the formation of a $^{22}$Ne-rich shell closer to the surface. In this case, a smaller cooling delay, which is actually required by current observational data, is generated. The fact that the {\it qualitative} outcome of $^{22}$Ne phase separation depends on the $^{22}$Ne mass fraction is the key attribute that allows this process to cause both small cooling delays at standard $^{22}$Ne abundances and very long cooling delays at moderately above-average $^{22}$Ne abundances. Detailed cooling sequences and population synthesis simulations will be needed to confirm our conclusions.

Future work should focus on building detailed evolutionary models to study the interplay of C/O phase separation and $^{22}$Ne settling in the liquid phase with $^{22}$Ne phase separation. The phase separation of $^{56}$Fe, another minor species with a potentially large impact on white dwarf cooling \citep{xu1992}, should also be investigated using modern simulation methods. Finally, this work motivates further efforts to better constrain the still poorly known C/O abundance profiles of white dwarfs \citep{giammichele2018}, whose shape strongly affects the amount of energy released by $^{22}$Ne phase separation.\\

We are grateful to the anonymous referee for useful comments that have improved the manuscript. S.B. also thanks A. B{\'e}dard and P. Brassard for useful discussions on the STELUM code. Research presented in this article was supported by the Laboratory Directed Research and Development program of Los Alamos National Laboratory under project number 20190624PRD2. This work was performed under the auspices of the U.S. Department of Energy under Contract No. 89233218CNA000001.

\bibliography{references}{}
\bibliographystyle{aasjournal}

\end{document}